# Selective advantage of aerobic glycolysis over oxidative phosphorylation

Alexei Vazquez

Nodes & Links Ltd, Salisbury House, Station Road, Cambridge CB1 2LA, England, UK

## Abstract

**The utilization of glycolysis in aerobic conditions have been a subject of debate for more than a century. A hypothesis supported by previous data is that glycolysis has a higher rate of ATP production per protein mass and per occupied volume than oxidative phosphorylation (OxPhos). However, a recent work by Shen et al[14] challenges previous estimates, reporting that OxPhos has a higher rate of ATP production per protein mass than glycolysis. Here I show that Shen et al[14] make a key assumption that is a subject of debate: that the proteomic cost of OxPhos is limited to proteins in enzymes of OxPhos and the TCA cycle. I argue that an intact mitochondria is required for functional OxPhos and therefore the whole mitochondrial protein content should be included for the cost estimate of OxPhos. After doing so, glycolysis is the most efficient pathway per protein mass or per volume fraction.**

When Usain Bolt crossed the finish line after running 100 meters in 9.58 seconds, his blood was full of lactate. His muscles were powered by glycolysis converting blood sugar to lactate and producing ATP.

Oxygenated cells ferment when they are operating close to their maximum metabolic rates, a phenomenon known as *aerobic fermentation* or *aerobic glycolysis* (~2ATP/glucose). It comes to a surprise. In those conditions cells could instead utilize oxidative phosphorylation (OxPhos, ~32 ATP/glucose). The consensus is that aerobic fermentation is either an empirical artifact or an anticipation of *hypoxic* conditions. Pasteur postulated that oxygen represses fermentation (*Pasteur effect*) and defined fermentation as "la vie sans gaz oxygene libre"[1]. Warburg discovered high rates of aerobic glycolysis in cancer cells (*Warburg effect*) and assumed mitochondrial defects[2]. Crabtree observed aerobic glycolysis in normal tissues infected by viruses and he postulated a repressive effect of high glycolysis rates on oxidative phosphorylation[3] (*Crabtree effect*). Yet, there is no empirical demonstration that oxygen limited the OxPhos activity in Usain Bolt's muscles.

Myself and others have proposed an alternative hypothesis: that in nutrient rich conditions and high metabolic rates the most efficient metabolic pathways are those maximizing metabolic rate per protein mass and per occupied volume[5-12]. In the context of energy production, that is the pathway maximizing ATP production rate per protein mass ($r_M$) and per volume ($r_V$) of the energysome. Previous estimates indicated that glycolysis has higher $r_M$ and $r_V$ than mitochondrial OxPhos[7,10,13] and therefore that glycolysis is more efficient in nutrient rich conditions and high metabolic rates.

This picture has been challenged by recent estimates by Shen et al[14]. According to their measurements+calculations, $r_M$ is actually smaller for glycolysis than mitochondrial OxPhos. However, there is an issue regarding what it is considered as OxPhos protein mass to calculate the proteomic cost. Shen et al assumed that the proteome mass associated with OxPhos is restricted to proteins in OxPhos enzymes and the TCA cycle. That assumption is a matter for debate. Based on yeast knockout libraries and siRNA screens in mammalian cells, about half of all mitochondrial related proteins are essential for optimal respiratory activity[15,16]. Therefore, an alternative assumption is that the proteomic cost of OxPhos is the whole mitochondrial protein mass, because the whole mitochondria needs to be present for OxPhos to function optimally.

The rate of ATP production per mitochondrial protein mass can be measured *in vitro* using isolated mitochondria. Table 1 reports a compilation of experimental reports for isolated yeast and mammalian mitochondria[17-26]. In addition, there is a report of muscle glycolysis reconstituted *in vitro*, that we can use as an independent estimate for glycolysis. The maximum value for OxPhos is 52 mmol ATP/h/g of mitochondrial protein. The maximum value for glycolysis is 88 mmol ATP/h/g of mitochondrial protein. Therefore, under the alternative assumption, that the proteomic cost of OxPhos is the whole mitochondrial protein mass, glycolysis has a 1.7 higher rate of ATP production per protein mass.

The advantage of glycolysis over OxPhos is more pronounced when focusing on the crowding cost. The specific volume of glycolysis can be estimated from the specific volume of proteins 0.73 ml/g[26]. For mitochondria we expect a larger specific volume. Since the intracellular media is composed of 30-40% macromolecules the mitochondrial specific volume should be in the range of 0.73 ml/g/0.3-0.4, that is 1.8-2.4 ml/g. The reported mitochondrial specific volume is 2.6 ml/g[27], close to the expected range. Normalizing by these specific volumes we obtain $r_V$, the rate of ATP synthesis per unit of energysome volume. The maximum value for OxPhos is 20 mmol ATP/h/ml of mitochondria. The maximum value for glycolysis is 120 mmol ATP/h/ml of glycolysisome. Therefore, under the alternative assumption, that the molecular crowding cost of OxPhos is the whole mitochondrial volume, glycolysis has 6 times a higher rate $r_V$ per occupied volume.

The massive effort by Shen et al[14] thus boils down to a key assumption, what makes the OxPhos associated proteome? We as a community should first agree on what proteins are accounted for when calculating the OxPhos rate per mass of protein. Shen et al[14] included proteins associated with OxPhos and TCA cycle reactions. I suggest the whole mitochondrial proteome, because OxPhos cannot function in the absence of intact mitochondria. It is also evident that while the rate per protein mass $r_M$ of fermentation is not far higher than that of OxPhos, the rate per occupied volume $r_V$ of fermentation is 5 times higher than that of OxPhos. This suggest that molecular crowding exerts a stronger selective pressure than proteome biosynthesis cost for the utilization of glycolysis at high metabolic rates. Molecular crowding is a metabolic constraint for both proliferating and non-proliferating cells and therefore applies more generally[28].

Table 1

| Energysome | Organism/Tissue | µmol ATP/min/mg | mmol ATP/h/g | mmol ATP/h/ml | Ref. |
|---|---|---|---|---|---|
| OxPhos | Rat liver | 0.15 | 9.0 | 3.5 | 17, Fig. 2 |
| | Rat liver | 0.2 | 12 | 4.6 | 18, Fig 12 |
| | Mouse liver | 1.1 | 66 | 25 | 19, Fig. 1A |
| | Human skeletal muscle | 0.35 | 21 | 8.1 | 20, Fig. 3 |
| | Mouse heart muscle | 0.4 | 24 | 9.2 | 21, Fig 6A |
| | Human skeletal muscle | 0.52 | 31 | 12 | 22, Table 2 |
| | Rat heart muscle | 0.5 | 30 | 12 | 17, Fig. 2 |
| | Rat skeletal muscle | 0.8 | 48 | 18 | 17, Fig. 2 |
| | Yeast | 0.866 | 52 | 20 | 23[a], Table 3 |
| | Yeast | 0.366 | 22 | 8.4 | 24, Table I |
| | Maximum | 1.1 | 66 | 25 | |
| Fermentation | Rat muscle 30°C | 0.575 | 35 | 47 | 25, Table 1 and 3 |
| | Rat muscle 40°C | 1.475 | 89 | 121 | 25, Table 1 and 3 |
| | Maximum | 1.475 | 89 | 121 | |

Table 1: Rates of ATP production per unit of protein mass or occupied volume. [a]Assuming ATP/O=1, as suggested for respiration rates above 300 natom O/min/mg protein (Ref. 23, Fig. 4).